\documentclass[prl,twocolumn,showpacs,floatfix,superscriptaddress]{revtex4}

\usepackage{hyperref}
\usepackage{graphicx}
\usepackage{amsmath}
\usepackage{amsfonts}

\newcommand{\mv}[1]{\left\langle#1\right\rangle}
\newcommand{\Heff}{H_\text{cc}}
\newcommand{\omegaB}{\omega_\text{B}}
\newcommand{\TB}{T_\text{B}}
\newcommand{\xB}{x_\text{B}}
\renewcommand{\Re}{\mathrm{Re}}
\renewcommand{\Im}{\mathrm{Im}}

\begin{document}

\title{Stable Bloch oscillations of cold atoms with time-dependent interaction}

\author{C.\ Gaul}
\affiliation{Physikalisches Institut, Universit\"{a}t Bayreuth,
D-95440 Bayreuth, Germany}

\author{R.\ P.\ A.\ Lima} 
\affiliation{GISC, Departamento de F\'{\i}sica de Materiales, Universidad
Complutense, E-28040 Madrid, Spain}

\author{E.\ D\'{\i}az}
\affiliation{GISC, Departamento de F\'{\i}sica de Materiales, Universidad
Complutense, E-28040 Madrid, Spain}

\author{C.\ A.\ M\"{u}ller}
\affiliation{Physikalisches Institut, Universit\"{a}t Bayreuth,
D-95440 Bayreuth, Germany}
\affiliation{Laboratoire Kastler-Brossel, UPMC, ENS, CNRS;
4 Place Jussieu, F-75005 Paris, France}

\author{F.\ Dom\'{\i}nguez-Adame}
\affiliation{GISC, Departamento de F\'{\i}sica de Materiales, Universidad
Complutense, E-28040 Madrid, Spain}

\begin{abstract}

We investigate Bloch oscillations of interacting cold atoms in a mean-field framework. 
In general, atom-atom interaction causes
dephasing and destroys Bloch oscillations. 
Here, we show that Bloch oscillations are persistent 
if the interaction is modulated harmonically with suitable frequency and phase. 
For other modulations, Bloch oscillations are rapidly damped.   
We explain this behavior in terms of collective coordinates whose Hamiltonian dynamics 
permits to predict a whole family of stable solutions. In order to describe    
also the unstable cases, we carry out a stability analysis  
for Bogoliubov excitations. Using Floquet theory, we are able to predict the unstable modes as well as their growth rate, found to be in excellent agreement with numerical simulations. 

\end{abstract}

\pacs{
03.75.Lm; % Tunneling, Josephson effect, Bose-Einstein condensates in periodic
          % potentials, solitons, vortices, and topological excitations 
52.35.Mw; % Nonlinear phenomena: waves, wave propagation, and other interactions 
          % (including parametric effects, mode coupling, etc.) 
37.10.Jk  % Atoms in optical lattices 
}

\maketitle

% Introduction

The dynamics of quantum 
particles in periodic potentials subjected to uniform acceleration
has a long history and fascinating physics. 
Electrons in tilted periodic potentials present dynamical localization and may undergo
coherent oscillations, both in real and in momentum space, known as Bloch
oscillations (BOs)~\cite{Bloch28}. 
BOs were observed for the
first time as coherent oscillations of electronic wave-packets in semiconductor
superlattices~\cite{Feldmann92,Leo92}. These oscillations are due to interference 
of partially scattered wave amplitudes and therefore can be observed for any
phase-coherent waves accelerated in periodic potentials. BOs have been 
directly observed with ultracold
atoms~\cite{BenDahan96,Wilkinson96}, Bose-Einstein condensates~\cite{Anderson98}
and noninteracting fermions~\cite{Roati04} in tilted optical lattices.

After excitation BOs persist until the quantum particles lose their phase coherence. 
In Bose-Einstein condensates interatomic interactions lead to a rapid broadening of the
momentum distribution and a strong dephasing. Even in the most favorable 
experimental conditions, only few cycles are usually observed.  It is then
believed that nonlinearities must generally lead to a breakdown of the 
BOs~\cite{Fallani04,Trombettoni01,Witthaut05}. 

\begin{figure}
\includegraphics[width=\linewidth]{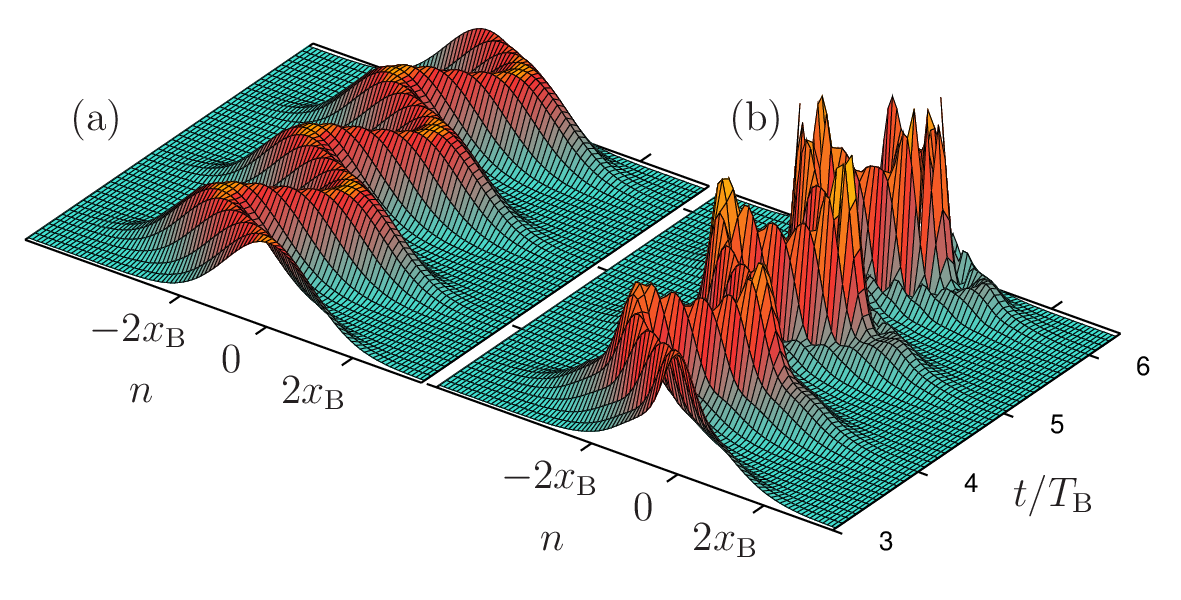}
\caption{Condensate density $|\Psi_n(t)|^2$ for two 
interactions modulated harmonically at the Bloch frequency $\omegaB=F$:  
(a)~$g(t)=g_0\cos(F t)$ and (b)~$g(t)=g_0\sin(F t)$ with $F=0.2$ and $g_0=1$.
Data from numerical integration of the Gross-Pitaevskii equation~\eqref{eqGPE}. 
The initial width of the wave packet is $\sigma_0=10$, equal to the amplitude $\xB=2/F$ of free BOs.  
While the BOs in (a) are perfectly stable, they are rapidly destroyed in (b). }
\label{fig1}
\end{figure}

In cold-atom experiments, one is able to change the
interaction strength by means of Feshbach resonances 
\cite{Donley01,Fattori08}.  
Reducing the interaction to zero, Gustavsson \emph{et al.}~\cite{Gustavsson08} 
were able to increase the dephasing time of a cloud of Cs atoms from a
value slightly larger than a Bloch period $\TB$ to more than $2\times 10^4\,\TB$.  
A different way of obtaining stable Bloch oscillations was proposed by Salerno \emph{et al.}~\cite{Salerno08}: 
properly
designing the spatial dependence of the scattering length around the zero
crossing in a Feshbach resonance should result in 
%long-loving 
long-living BO of bright solitons.   
In this Letter, we propose instead 
to modulate the interaction harmonically in time by an oscillating, but spatially homogeneous 
magnetic field close to a Feshbach resonance \cite{Abdullaev03}, which should be much easier experimentally.
Although in general atom-atom interactions lead to a rapid dephasing
of BOs, 
remarkably with some harmonic modulations we find
stable oscillations of the condensate [see Fig.~\ref{fig1}(a)], while for 
others the oscillations are rapidly dephased [see Fig.~\ref{fig1}(b)].

%--------------------
%--------------------

Our starting point for this study  is the mean-field 
Gross-Pitaevskii equation for the discrete tight-binding model~\cite{Morsch06} 
\begin{equation}
i \dot \Psi_n = -\Psi_{n+1}-\Psi_{n-1} + F n \Psi_{n} + 
g(t) |\Psi_{n}|^2 \Psi_{n} \, ,
\label{eqGPE}
\end{equation}
where $\Psi_n$ denotes the wave function at each potential well $n$ of the optical lattice.
The overdot indicates the time derivative, the hopping parameter is set as the unit of energy, and $\hbar=1$. 
We assume that initially the
atomic cloud has a Gaussian density profile centered at rest in the 
lattice, 
$\Psi_n(0)=(2\pi\sigma_{0}^2)^{-1/4}\,\exp\left[-n^2/4\sigma_{0}^2\right]$.
In the noninteracting case $g(t)=0$, the atomic cloud oscillates with the
Bloch frequency $\omegaB=2\pi/\TB=F$. 

To investigate the interacting case, we first solve (\ref{eqGPE}) 
numerically by means of the fourth-order Runge-Kutta method. 
Figure~\ref{fig1} shows the time evolution of 
the condensate density $|\Psi_n(t)|^2$ 
for two different harmonic
interactions modulated at the Bloch frequency. 
For $g(t)=g_0\cos(F t)$, 
we observe that the initial Gaussian shape is preserved over time, and the
condensate performs perfectly stable BOs with frequency $\omegaB$. 
On the contrary, for $g(t)=g_0\sin(F t)$, the initial shape is distorted already after few Bloch cycles. As a
consequence, BOs are rapidly damped.

%--------------------
%Collective coordinates

In order to understand how the interaction affects BOs of a Gaussian wave packet, we start with a description in terms of \emph{collective coordinates} 
\cite{Trombettoni01}.  As free variables, we chose the center-of-mass position 
$x(t) = \mv{n}= \sum_n n |\Psi_n(t)|^2$ and width $w(t) =  \mv{(n-x(t))^2}$ and their conjugate momenta $p(t)$ and $b(t)$, respectively. 
These are defined by their generating r\^ole, $-i\partial_p\Psi_n= n \Psi_n$ and similarly for $b$. 
The wave function is up to a global phase parametrized as 
\begin{equation}\label{psiA.eq}
\Psi_n(t)=e^{i p(t)n} A(n-x(t),w(t),b(t)), 
\end{equation}
where the rapidly varying exponential $\exp[ip(t)n]$ is factorized from a 
smooth Gaussian envelope 
\begin{equation}\label{envelope.eq}
A(n,w,b) = \frac{1}{(2\pi w)^{1/4}} 
\exp\left[-\frac{n^2}{4w}+ib n^{2} \right].
\end{equation}
The equation of motion (\ref{eqGPE}) derives as $i\dot\Psi_n = \partial H/\partial\Psi_n^\ast$  
from the nonlinear Hamiltonian  
\begin{equation}\label{Hamiltonian.eq}
H = \sum_n \left\{-(\Psi_{n+1}\Psi_n^\ast+c.c.) + F n |\Psi_n|^2 + \frac{g(t)}{2}|\Psi_n|^4 \right\}. 
\end{equation}
Inserting the ansatz (\ref{psiA.eq}), Taylor-expanding the discrete gradient to second order, and performing the Gaussian integration, we find the effective Hamiltonian 
\begin{equation}\label{Heff.eq}
\Heff  = Fx - \left(2-\frac{1+16b^2w^2}{4w}\right)  \cos p+ \frac{g(t)}{4\sqrt{\pi}}w^{-1/2}.
\end{equation}
By construction, the collective coordinates  obey the canonical equations of motion 
\begin{subequations}
\begin{eqnarray}
\dot p &=&- \frac{\partial \Heff}{\partial x} = -F,    \label{pdot} \\
\dot x &=& \frac{\partial \Heff}{\partial p} = \left(2-\frac{1+16b^2w^2}{4w}\right)  \sin p, \label{xdot}\\
\dot b &=& - \frac{\partial \Heff}{\partial w} = \frac{1-16w^2b^2}{4w^2}\cos p +  \frac{g(t)}{8\sqrt{\pi}} w^{-3/2}, \label{bdot} \\    
\dot w &=& \frac{\partial \Heff}{\partial b} = 8 w b \cos p.   \label{wdot}  
\end{eqnarray}\label{eq:cvcem}
\end{subequations}
The initial conditions for the case under study are $x(0)= 0$, $p(0)=0$, $w(0) = \sigma_0^2$, $b(0)=0$.   
Equation (\ref{eq:cvcem}) shows that within this ansatz, the momentum of the centroid is always given as $p(t) = -Ft$ which then serves as the driving term in the other equations. The autonomous equations (\ref{bdot}) and (\ref{wdot}) describing the width can be solved first, their solution finally conditioning the centroid motion obtained by integrating (\ref{xdot}).  

Let us study the implications of these equations in some simple cases. 
The complex width of the Gaussian envelope, $z = [w^{-1}-4ib]^{-1}$, evolves according to 
\begin{equation}\label{zdot.eq}
\dot z = i \cos Ft + \frac{ig(t)}{2\sqrt{\pi}} \frac{z^2}{|z|^3}(\Re z)^{3/2} . 
\end{equation} 
In the linear case $g=0$, this equation has the exact solution $\Re z_0 = w(0) = \sigma_0^2$ and $\Im z_0(t) = F^{-1}\sin Ft$. It describes a Gaussian wave packet with a breathing width 
$w_0(t) = \sigma_0^2\left[ 1+ (\sin Ft)^2/(F \sigma_0^2)^2\right].$  
This solution is valid for small $1/(F\sigma_0^2)$, i.e., sufficiently broad wave packets. 
Within the collective-variable approach, the amplitude of BOs  
is actually determined by $R:= (8\Re z)^{-1} = (1+16 b^2 w^2)/(8w)$, appearing in (\ref{xdot}). 
In the linear case, its constant value $R_0= \sigma_0^{-2}/8\ll 1$ yields the usual BOs $x_0(t) = \xB \cos Ft$ with amplitude $\xB=2(1-R_0)/F$. 

A constant nonlinearity $g(t)=g_0$ leads to a damping of the oscillation amplitude. 
With the highly nonlinear and rapidly oscillating equations of motion (\ref{bdot}) and (\ref{wdot}) one finds $\ddot{R} = g_0 \mathcal{W}(t) \cos Ft + g_0^2/(16\pi w(t)^2)$. The first term contains contributions from $w(t)$ and $b(t)$ that are very effectively suppressed  over a Bloch cycle by the oscillating factor $\cos Ft$. The decay to lowest order in $g_0$ therefore is driven by the second term, and the amplitude decreases initially like 
$\xB(t) = (2/F)(1- R_0 - 2 g_0^2R_0^2t^2/\pi)$
\cite{Trombettoni01,Witthaut05}.

Let us now consider a time-dependent interaction of the form $g(t)=g_0\sin(\omega t+\delta)$.  
The equations of motion for the collective coordinates %~(\ref{EOM}) 
allow to identify a whole family of values $(\omega,\delta) $ that result in 
strictly periodic motion. 
Formally, we can combine (\ref{bdot}) and (\ref{wdot}) to a single equation $\dot v= X(v,t)$ for $v=(w,b)$. 
This Hamiltonian flow is driven by a vector field $X(\cdot,t)$ that depends explicitly on $\cos Ft$ and $g(t)$.  
Let now $\omega= (\nu_1/\nu_2)\omegaB$ with $\nu_1, \nu_2 \in \mathbb{N}$ be commensurate  with the Bloch frequency $\omegaB=F$.  Then $X(\cdot,t+T) = X(\cdot,t)$ is periodic with $T=\nu_2 \TB$.
Since different trajectories $v(t)$ cannot intersect, the flow dynamics is strictly periodic if there exists a time $\tau$ such that $v(\tau-T/2) = v(\tau+T/2)$. 
This in turn is guaranteed by the equation of motion if $X(\cdot,t)$ is odd around $\tau$, $X(\cdot,\tau-t) = -X(\cdot,\tau+t)$. But since $X$ contains $\cos Ft$, this time can only be one of its zeroes 
$\tau_j=(2j+1)\frac{\pi}{2}F^{-1}, j\in\mathbb{N}$. 
This then requires also the interaction to be odd at that point, which fixes $\delta_j=-\omega \tau_j$. 
Therefore, the family of time-dependent interactions with periodic solutions contains all linear combinations of
\begin{equation}
\label{stableg.eq}
g(t) = g_0 \sin\left(\frac{\nu_1}{\nu_2}[Ft - \frac{\pi}{2}  
(2j+1)] \right), \quad \nu_1,\nu_2,j \in \mathbb{N} \, , 
\end{equation}
for example $\cos\left((2n+1) F t\right)$, $\sin(2n F t)$, $\cos(F t/3)$ or $\sin(2F t/3)$.  
In all such cases, the periodicity of $w(t)$ and $b(t)$ predicts, via (\ref{xdot}), perfect BOs in spite of the underlying nonlinearity.

%added paragraph
This stabilization mechanism does not rely on 
suppressing the nonlinearity at the band edges, which can in general be helpful to avoid Landau-Zener tunneling 
\cite{Morsch06} but is irrelevant in our single-band description. 
Instead, the simplest stable interaction of type $g(t) = \pm \cos(Ft) = \pm \cos(p(t))$ 
has its maximum amplitude with positive (negative) sign at the
band center $p(t) = 0$ and negative (positive) sign at the band edge
$p(t) = \pi$. Clearly, it does not vanish at the band edge, but rather at the band 
midpoint, where the mass changes sign and \emph{dynamical instabilities} would start to appear
\cite{Fallani04,Wu03}. 
%

%--------------------
\begin{figure}
\centerline{\includegraphics[width=80mm,clip]{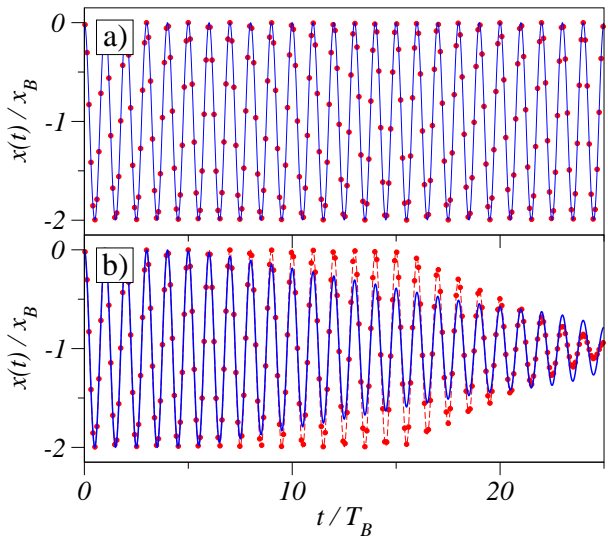}}
\caption{Centroid $x(t)$ for (a) $g(t)=g_0\cos(\omega t)$,
(b) $g(t)=g_0\sin(\omega t)$ with $\omega=F=0.2$ and $g_0=1.0$.
The solid line is obtained by numerical integration of \eqref{eqGPE}; 
circles connected by dashed lines are the solution of the collective-coordinate Eqs.~\eqref{pdot}-\eqref{bdot}; 
In the stable case (a) both curves agree, whereas in (b) the collective coordinates predict a later onset of damping.
}
\label{fig2}
\end{figure}
%--------------------

We have 
tested the collective-variable predictions by comparing the centroid $x(t)$ 
as obtained from equation~(\ref{eq:cvcem}) with the results from the
numerical integration of~(\ref{eqGPE}).
Collective coordinates provide
an excellent description of long-living BOs [Fig.~\ref{fig2}(a)]. 
%In the damped case $g(t)=g_0\sin(F t)$ this approach does also predict the
%breakdown of BOs. However, it overestimates the time until the onset of damping,
%as seen in Fig.~\ref{fig2}(b). Within the ansatz (\ref{psiA.eq})
%the width of the wavepacket is the only degree of freedom that can respond to the nonlinear perturbation. 
%A closer look at Fig.~\ref{fig1}(b) reveals that it follows correctly the contraction of the wavepacket, but by construction cannot deviate from the Gaussian shape.
%Thus this ansatz misses the actual dephasing mechanism, which involves the excitation of fluctuations. 
%%%%replaced by:
In the damped case $g(t)=g_0\sin(F t)$ this approach also predicts a 
breakdown of BOs, but not quite correctly [Fig.~\ref{fig2}(b)]. 
This ansatz allows only the gaussian width to respond to the nonlinear perturbation, but it cannot capture 
the actual dephasing mechanism, which involves the excitation of fluctuations as apparent from Fig.~\ref{fig1}(b).  
%--------------------

%--------------------
% Beyond the collective coordinates approach

To proceed \emph{beyond collective coordinates} we start with an
infinitely wide wave packet, namely an atomic cloud with uniform density $|\Psi_{n}^{0}|^2=\rho_0$. 
In this case, equation \eqref{eqGPE} is exactly solved by 
$\Psi_{n}^{0}(t)=\sqrt{\rho_0}\,\exp[-i\phi(t)]$ with 
\begin{equation}
\phi(t)= Fnt -\frac{2}{F}\,\sin (F t)+\int_{0}^{t}d t^{\prime}\mu(t^{\prime}) 
\label{solUNIFORM}
\end{equation}
in terms of the chemical potential $\mu(t)=\rho_0g(t)$. This solution predicts that an infinitely narrow momentum distribution will perfectly Bloch oscillate in momentum space, i.e., the condensate current will oscillate homogeneously without signatures in position space, and this for 
arbitrary interaction.  

In the following we study the broadening of the momentum distribution due to the 
\emph{growth of small perturbations}. 
If the wave function deviates slightly from the homogenous solution, 
$\Psi_n(t)=\left[\sqrt{\rho_0}+\Phi(n,t)\right]\exp[-i\varphi(t)]$, 
linearizing  (\ref{eqGPE}) gives the equation of motion for the fluctuation $\Phi(n,t)$:     
\begin{eqnarray}
i \dot {\Phi} &=& -  \cos (Ft) \Phi''
+ 2i \sin (Ft) \Phi'  + 2 \mu(t) \Re\Phi\,. 
\label{eqDeltaPsi}
\end{eqnarray}
$\Phi(n,t)$ is assumed to be a smooth function with spatial derivative $\Phi'=\partial_n\Phi$.  
The interaction term acts on $\Re\Phi$, i.e., the
density fluctuation.
Separating real and imaginary part, 
$\Phi(n,t) = s(m,t) + i\,d(m,t)$ while 
transforming to the reference frame $m=n-x_0(t)$ of the unperturbed BO with $x_0(t) = 2F^{-1}\cos Ft$, and going to the momentum representation brings \eqref{eqDeltaPsi} to
\begin{subequations}
\begin{eqnarray}
\label{eqStabilityproblem}
\dot d_k &=& -\left[ k^2 \cos(F t) +2\mu(t)\right]  s_k\, , \\
\dot s_k &=& k^2\cos(F t) d_k\, .
\end{eqnarray}\label{eqStability}%
\end{subequations}
These are the Bogoliubov-de Gennes equations for excitations
of a homogeneous condensate with chemical potential $\mu(t)$
and a time-dependent mass, such that $\epsilon_k^0(t) = k^2 \cos(F t)$. 
The width of the momentum-space distribution,   
$(\Delta k)^2 = (N\rho_0)^{-1}\sum_k (k-p)^2 |\Psi_k|^2
 = (N \rho_0)^{-1} \sum_k k^2 \big(|d_k|^2 + |s_k|^2\big) 
$, is given by these amplitudes.  
Thus, their stability is the key to understand the stability of BOs.

In Eqs.~\eqref{eqStability}, we have again the structure of a first-order Hamiltonian equation of motion (now with a linear evolution operator) depending on $\cos(F t)$ and $\mu(t)=\rho_0g(t)$. Thus the same argumentation as for  
the collective coordinates holds: solutions are periodic and hence perturbations do not grow if $g(t)$ is of the type  \eqref{stableg.eq}. 

But now we are also able to make quantitative predictions about both stable and unstable cases. Thanks to the linearity of \eqref{eqStability}, 
\emph{Floquet theory}~\cite{Teschl08} 
applies 
if the interaction is modulated at a commensurate frequency $\omega = (\nu_1/\nu_2)\omegaB$. 
The stability of this system can be assessed by integrating the differential
equations~(\ref{eqStability}) over one period $T=\nu_2 \TB$ with two different initial
conditions: $\{d_k^{1}(0)=1,s_k^{1}(0)=0\}$ and
$\{d_k^{2}(0)=0,s_k^{2}(0)=1\}$. 
A perturbation with wave vector $k$ is stable when
$|\Delta_k|\leq1$, where $\Delta_k=(1/2)\left[ d_k^{1}(T)+s_k^{2}(T)\right]$.
In the periodic cases $\Delta_k=1$.
An unstable fluctuation with wave vector $k$ grows exponentially with the Lyapunov exponent 
$\lambda_k = T^{-1} \log\left[\Delta_k + \sqrt{\Delta_k^2 - 1}\right]$.

%--------------------
\begin{figure}%[ht]
\centerline{\includegraphics%[height=55mm,width=60mm,clip]
{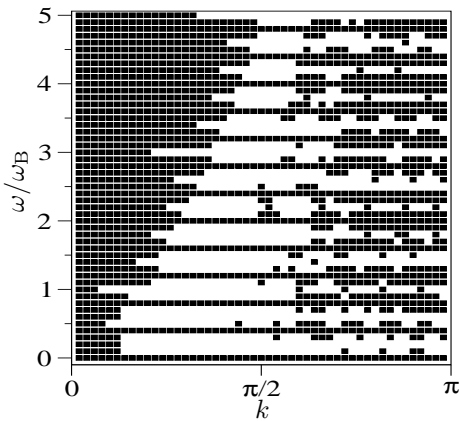}}
\caption{Stability map for $g(t)=g_0\sin(\omega t)$, with $\omegaB=F=0.2$ and $g_0=1.0$. 
Black squares indicate the wave vectors $k$ of stable excitations at each frequency $\omega$. 
The stable cases at $\omega/\omegaB = 2n/5$ predicted by \eqref{stableg.eq} appear as full lines; the other stable lines are not resolved due to the finite frequency step $0.1$ used here.  
For the case of $\cos(\omega t)$, the diagram is similar, with stable lines at $\omega/\omegaB = (2n+1)/5$.
}
\label{fig3}
\end{figure}
%--------------------
 
Figure~\ref{fig3} indicates by black squares those parameters $(\omega/\omegaB,k)$ in the case $g(t)=g_0\sin(\omega t)$, for which $|\Delta_k|\leq1$ assures stability. 
Full lines correspond to the stable cases at $\omega/\omegaB = 2n/5$, $n\in \mathbb{N}$, predicted by \eqref{stableg.eq}; other stable lines are not resolved due to the finite frequency resolution $0.1$ used for the plot.  
For most frequencies, there are many $k$-vectors belonging to unstable excitations, so the BO will decay  [Fig.~\ref{fig1}~(b)]. 

\begin{figure}
\includegraphics[height=0.95\linewidth,angle=270]{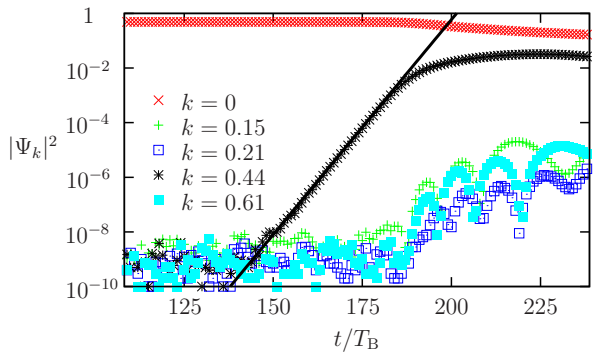}
\caption{Momentum density $|\Psi_k|^2$ for selected $k$-modes in the unstable case $g(t) = g_0 \sin(F t)$. 
The original wave function is centered around $k=0$. The solid line marks the growth rate of the most unstable mode as predicted by Floquet theory. 
The growth of this mode precedes the damping of the centroid motion that sets in at $t\approx 200 \TB$. 
Numerical parameters: $\sigma_0=100$, $\mu_0 = 0.01$, $F=0.2$. 
}\label{figWideK3D}
\end{figure}

For a quantitative test of Floquet-theory predictions, we choose a larger width $\sigma_0=100$ and a weak interaction parameter $\mu_0 = 0.01$.
As obvious from Fig.~\ref{fig3}, unstable modes are typically encountered at intermediate values of $k$. 
Floquet analysis predicts as the most unstable mode $k\approx 0.44$ with a Lyapunov exponent of 
$\lambda_\text{max}\approx 0.1815\,\TB^{-1}$. 
The numerically calculated centroid motion (similarly to Fig.~\ref{fig2}) starts to get damped around 200$\TB$. The $k$-space picture, Fig.~\ref{figWideK3D}, is much more revealing and shows that the most unstable mode starts to grow much earlier. The numerically observed dominant mode and also its growth rate are in perfect agreement with our Floquet-theory predictions. 

%--------------------
%--------------------

To conclude, we have shown that stable BOs of cold atoms are possible
when the  atom-atom interaction is modulated in time with suitable frequency and phase. 
Collective coordinates are found to accurately describe undamped BOs. For commensurate frequencies $\omega$ and $\omegaB$, we have identified a class of time-dependent interactions, for which the BOs remain perfectly periodic.
But collective coordinates cannot capture the dynamics of unstable cases where the main wave packet decays by radiating excitations.  
In order to explain the dephasing of the oscillations in unstable cases, 
we applied a stability analysis based on Floquet theory that is in excellent agreement with the numerical results. 
Using harmonically modulated interaction thus opens the possibility of studying the precise influence of other dephasing mechanisms on Bloch oscillations, such as deliberately introduced disorder
\cite{Drenkelforth08}. 

%--------------------
\begin{acknowledgments}
Travel between Bayreuth and Madrid is supported by the DAAD-MEC joint program
\emph{Acciones Integradas}. Work at Madrid was supported by MEC (Project
MOSAICO) and BSCH-UCM (Project PR34/07-15916). R.\ P.\ A.\ L.\ acknowledges 
support by MEC through the \emph{Juan de la Cierva\/} program and G. Rowlands
for helpfull discussions. C. G. acknowledges support by DFG and DAAD, and C.\ M.\ 
acknowledges helpful discussions with Y. Gaididei.  
 
\end{acknowledgments}

\end{document}